\documentclass[preprint,twoside,letterpaper,showpacs,amsmath,amssymb,aps]{revtex4}
\usepackage{graphicx}% Include figure files
\usepackage{dcolumn}% Align table columns on decimal point
\usepackage{bm}% bold math
\usepackage{color}
\usepackage{textcomp}
\usepackage[pdfpagelayout=TwoColumnLeft,bookmarksopen,bookmarksopenlevel=1,pdfstartview=FitH]{hyperref}

\begin{document}

\newcommand{\lunibc}{LuNiBC \/}
\newcommand{\luo}{Lu$_{\mbox{\scriptsize 2}}$O$_{\mbox{\scriptsize 3}}$ \/}
\newcommand{\ro}{\textit{R}$_{\mbox{\scriptsize 2}}$O$_{\mbox{\scriptsize 3}}$ \/}
\newcommand{\luni}{LuNi$_{\mbox{\scriptsize 2}}$B$_{\mbox{\scriptsize 2}}$C \/}
\newcommand{\nib}{Ni$_{\mbox{\scriptsize 2}}$B \/}
\newcommand{\rni}{\textit{R}Ni$_{\mbox{\scriptsize 2}}$B$_{\mbox{\scriptsize 2}}$C \/}
\newcommand{\honi}{HoNi$_{\mbox{\scriptsize 2}}$B$_{\mbox{\scriptsize 2}}$C \/}
\newcommand{\yni}{YNi$_{\mbox{\scriptsize 2}}$B$_{\mbox{\scriptsize 2}}$C \/}
\newcommand{\alo}{Al$_{\mbox{\scriptsize 2}}$O$_{\mbox{\scriptsize 3}}$ \/}
\newcommand{\yo}{Y$_{\mbox{\scriptsize 2}}$O$_{\mbox{\scriptsize 3}}$ \/}
\newcommand{\sto}{SrTiO$_{\mbox{\scriptsize 3}}$ \/}
\newcommand{\tc}{T$_{\mbox{\scriptsize c}}$}
\newcommand{\hc}{H$_{\mbox{\scriptsize c2}}$}

\title[Growth and characterization of epitaxial \luni thin films on MgO single crystals]
{Growth and characterization of epitaxial \luni thin films on MgO single
crystals}

%\author{T. Niemeier$^{1,2,}$\footnote{Email: t.niemeier@ifw-dresden.de}}
\author{T. Niemeier\footnote{Email: tim.niemeier@ifw-dresden.de}, R. H\"uhne, A. K\"ohler, G. Behr, L. Schultz and B. Holzapfel}

%\thanks{\emph{Present address:} Insert the address here if needed}
%}                     % Do not remove
%\affiliation{$^1$IFW Dresden, P.O. Box 270116,D-01171 Dresden, Germany}
\affiliation{IFW Dresden, P.O. Box 270116, D-01171 Dresden, Germany}
\date{Received: \today}
% / Revised version: date}
% The correct dates will be entered by Publishing Company
%
\begin{abstract}
Thin films of \luni have been prepared on MgO(110) single crystal substrates using standard pulsed laser deposition from a stoichiometric target. Via the precise control of substrate temperature, laser energy density and pulse ablation rate, perfect epitaxial c--axis growth of the \luni phase was reproducibly achieved with an in--plane alignment of about 1 \textdegree\/ and an out--of--plane order of about 2.5 \textdegree. The samples prepared under optimized deposition conditions obtain critical temperatures of up to 15.8 K and steep superconducting transitions of about 0.3 K in the best samples. Residual resistivity ratios up to 15 were measured in the unstructured samples. The upper critical field $H_{c2}$ was resistively measured along the [001] direction on a sample and fitted with a power law. The value of $H_{c2}(0)$ = 9.82 T obtained from that fit is in good agreement with single crystal data and the power exponent describing the positive curvature for small external magnetic fields $\alpha$ = 0.19 indicates a relatively low intraband scattering in the films.
\end{abstract}
\maketitle \nopagebreak

\section{Introduction}
Since their discovery in 1994 \cite{CZB+94,CBS+94,SZC+94,SCKP94}, superconducting rare earth nickel borocarbides \rni [\textit{R} $\epsilon$ \{Y, Dy \ldots Tm, Lu\}] have been extensively investigated for their structural, magnetic and superconductive properties. In particular, the interaction between rare earth magnetism and superconductivity has been studied in detail. The results have been published in several extensive review articles \cite{MN01,TZ05,MN05,MSFD08}. During the years, all those  compounds have then been synthesized as single crystals using different methods. Traditionally, small plate-shaped crystals of the highest quality were grown from a \nib flux \cite{Mat95,BC06}. Recently, the increasing experience on the growth of these materials using a travelling floating zone technique (TFZ) and a subsequent thermal treatment resulted in much larger crystals of also very high quality \cite{SBKL05,BLS+08}. Therefore, the precise investigation of properties of this class of materials such as thermodynamical, superconductive or magnetic properties, which are highly dependant on the sample purity, became again an interesting field of research (see for examples: \cite{BIB+07,NBY+07,BCN+08,SGK+08}.
Since that time, there was much interest to grow textured thin films of these materials using epitaxial growth. Different groups successfully prepared borocarbide thin films and studied their structural, intrinsic and extrinsic properties \cite{AHNT98,AAC+97,ACG+01,GMP+01,HHHS00,WHSH01,WSH03,WSH04,BNY+05}. All those films were fabricated using physical vapor deposition techniques (PVD) on unbuffered or metal buffered oxidic single crystal substrates.
Basically, several specific challenges have to be met for the preparation of intermetallic Borocarbide thin films: 1.) High crystallization temperature: The crystalline bulk phase formation temperature is about 1550 \textcelsius. While the first films were produced by room temperature sputtering and annealed afterwords\cite{AHH+94}, a successful \textit{in situ} preparation of epitaxial c--axis--oriented borocarbide thin films was not reported for growth temperatures below 700 \textcelsius. 2.) High sensitivity of the rare earth material against oxidation: Oxygen incorporation into the borocarbide phase has to be prevented during thin film growth. As a first consequence, ultra high vacuum (UHV) conditions are necessary for physical vapor deposition. Secondly, oxygen interdiffusion into the film must be either hindered choosing suitable buffer materials or has to be at least controlled in a way that enables clean borocarbide phase growth on top. 3.) High diffusion of metallic buffer materials: The use of refractory metal buffers such as tungsten can suppress the oxidation of the rare earth but at the same time, interface reactions resulting in compounds as for example tungsten carbide may lead to changes of the composition of the borocarbide phase \cite{HHHS00,GMP+01}.
To state the challenge more precisely, no stable buffer or substrate has been found that would allow a direct growth of borocarbides without a significant metallic or oxidic interdiffusion at all. While trying metallic buffers as well, most groups eventually used oxidic substrates such as magnesium oxide (MgO) or sapphire single crystals as those are chemically stable up to high temperatures. During their experiments, it overall turned out that at the interface between borocarbide and oxidic substrate, a \ro layer is formed during the deposition of the \rni material. The thickness of this oxide interface layer, whose formation can be hindered by suitable metallic buffer layers as already mentioned\cite{HHHS00,GMP+01}, can reach about 100 nm\cite{CSS04}. Moreover, epitaxial growth can be transferred from the substrate through the interfacial oxide layer leading to a textured \rni phase under suitable deposition conditions. Investigations using transmission electron microscopy (TEM) revealed in detail how the in--plane alignment is transferred from the substrate to grains of the oxide interface layer and further to the borocarbide layer and located additional secondary phases at the substrate interface \cite{RWHK02,CSS04}. The growth control of this buffer layer is therefore one of the key parameters for the successful preparation of epitaxial \rni films.
However, the availability of epitaxial \rni films with appealing superconducting properties is rather limited. Various reports are present about the preparation of borocarbide thin films by sputtering or pulsed laser deposition (PLD), explicitly aiming for well--aligned textured \rni layers, high critical temperatures close to bulk values and high residual resistivity ratios, but achieving sufficient properties reproducibly at the same time has been reported only rarely. Mainly, two groups reported on significant progress in growing highly textured as well as in--plane ordered borocarbide thin films with good superconductive properties at the same time \cite{GMP+01,WHSH01}.
In this report, the preparation of epitaxial \luni thin films is presented starting from the target production until the final thin film samples. \luni was chosen because it is a non magnetic borocarbide and has the highest critical temperature of all stable rare earth nickel borocarbides. Single crystals of \luni show a superconducting transition temperature of up to 16.5 K \cite{MWK+97}. MgO(110) substrates are used which can lead to an improved in--plane texture in comparison to MgO(100) substrates\cite{FGB+03}. The growth parameters are therefore adapted for the use of these substrates enabling the reproducible synthesis of epitaxial superconducting \luni thin films showing critical temperatures close to 16 K and residual resistivity ratios up to 15 in the as-prepared state.
The paper is structured as follows: First, the sample preparation is explained in detail. The structural investigations using x--ray diffraction in standard Bragg--Brentano geometry as well as texture measurements are presented in the next part. Finally, the  superconducting properties of a representative thin film sample are shown and analyzed.

\section{Experimental}
\subsection{Target preparation}
The target preparation is the similar to the feed rod preparation for TFZ single crystal growth \cite{BLS+08}: For the target, lutetium swarfs of high purity was cut into small pieces (weight approximately $<$ 500 mg each) under argon atmosphere and cold-pressed at 30 kN together with a mixture of carbon (99.9+ wt\%) and nickel powder (99.99 wt\%, both from \emph{MaTeck}) and boron powder (99.52 wt\% from \emph{Eagle Picher}) in a stoichiometric 1:2:2:1 ratio. The received pill was molten several times in an induction melting oven under argon atmosphere (background pressure $< 2 \cdot \rm{10^{-5}}$ mbar) at about 1550 \textcelsius\/ and homogenized for a few minutes at about 1100 \textcelsius. The molten pill was cooled down and again molten two times (once from each side) in an arc discharge oven under argon atmosphere to achieve an oblate shape of approximately 20--25 mm diameter. Rapid cooling on a water-cooled copper plate was used to suppress large grain growth. Finally, the target was laterally cut with a water-cooled wire saw and slightly grinded to its necessary shape. After sawing, some small lunkers (diameter roughly smaller than 1 mm) were observed in the center of the target, which were removed by grinding as good as possible. Pictures of selected steps of the target production process are shown in Fig.\,1.

\subsection{Sample preparation}
Thin film deposition took place in a standard ultra high vacuum chamber with a base pressure of about 2 $\cdot \rm{10^{-9}}$ mbar. The laser ablation was performed with a KrF laser (\emph{Lambda Physik LPX 305}) at a wave length of 248 nm. The laser spot size was set to 1.5 x 4 mm approximately and the laser energy per pulse was chosen at 250 mJ giving an energy densityof about 4 J/cm². 
\begin{figure}[t!]
\includegraphics[width=7cm]{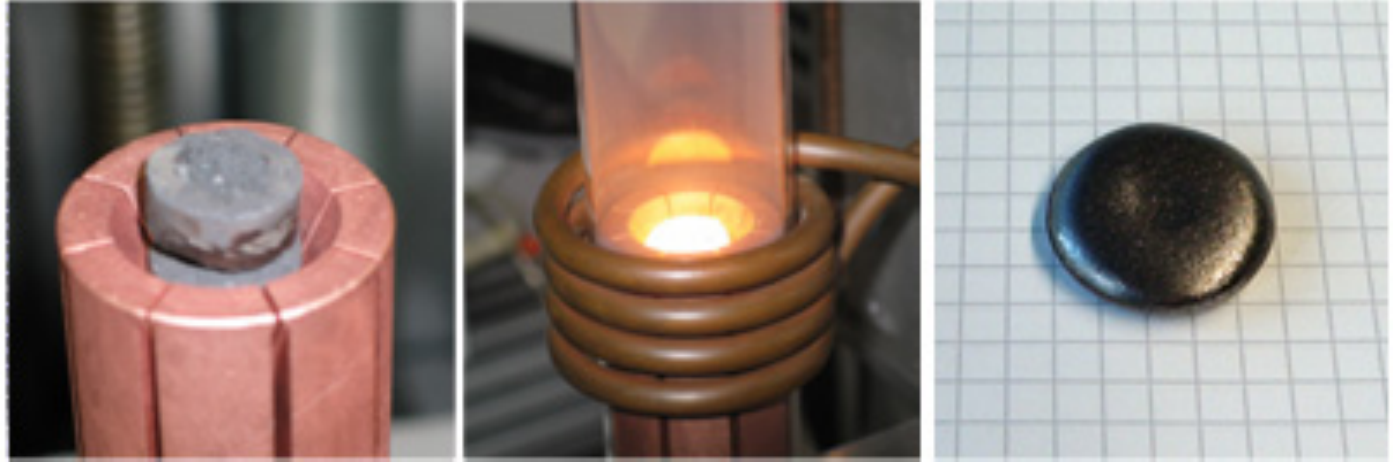}
\caption{(color online.) Pictures of selected steps of the target preparation: (a) pressed pill, (b) induction melting and (c) target after last arc discharge melting step.}
\end{figure}
A deposition rate of 0.005--0.01 nm per laser pulse was measured with an \emph{Inficon} rate monitor depending on the laser fluence and the target roughness using a target--substrate distance of 45 mm. The total film thickness $d$ was intended to 200 nm and the laser repetition rate $f$ was chosen between 5 and 20 Hz. Prior to thin film deposition, one--side polished, single--crystalline and [110]--edge--oriented MgO substrates with a size of 10 x 10 x 1 mm³ (from \emph{Crystec}) were ultrasonically cleaned in acetone for 5 min, then transferred to the chamber and heated \textit{in situ} under vacuum at about 1050 \textcelsius\/ for 5 minutes. The substrates were then cooled down to the deposition temperature according to sample and held for 15 minutes in order to stabilize the surface temperature. The target was cleaned before the deposition using a number of additional laser pulses while the substrate was protected by a shutter. Furthermore, the substrate holder was axially slowly rotated during the deposition to improve the homogeneity of the growing films.
Temperatures were measured using an \emph{Infrawin} infrared pyrometer before and directly after deposition of the film. The films were analyzed with x--ray diffractometry (\emph{Philips X'Pert} system) in Bragg-Brentano geometry using Co K$_{\rm{\alpha}}$ (standard Bragg-Brentano) or Cu K$_{\rm{\alpha}}$ (texture measurements) radiation. Superconductive properties of the films (patterning was not applied) were determined using resistive measurements in a \emph{Quantum Design PPMS}.

\section{Phase formation}
A series of\luni thin films has been grown on MgO(110) single crystal substrates at different deposition temperatures as described. The laser repetition rate was fixed at 20 Hz for this series. The phase formation, analyzed using standard Bragg Brentano x--ray diffraction, is shown in Fig.\,2 in dependence of the post--deposition temperature. We draw three main conclusions from the data: (1) The patterns show that the borocarbide 1221--phase is formed in a pronounced c--axis oriented growth for the whole range of the investigated deposition temperatures. The c--axis lattice parameters obtained from the patterns fit well with the data obtained by powder diffractometry. For higher temperatures, the amount of the c--axis oriented \luni phase is considerably increased. All in all, the c--axis texture of the borocarbide phase is almost perfect. (2) A perfectly [110]--textured \luo phase is observed in the measurements as well. This oxide is most likely formed at the interface via a rare earth oxidation reaction with the substrate as described ny other groups \cite{GMP+01,CSS04}. (3) For high temperatures, an increasing amount of secondary phases is formed indicating strongly increased diffusion reactions. The most pronounced peak found at $2\theta$ = 49.8\textdegree\/ can be addressed to a slightly off-stoichiometric \nib phase, which is likely formed during the lutetium oxidation at the substrate interface. A similar phase was was already found in earlier studies on \yni films \cite{HHHS00}. Based on those results, it is assumed that this impurity phase is most probably located within the interface region as well.
\begin{figure}[h]
\begin{center}
\includegraphics[width=8cm]{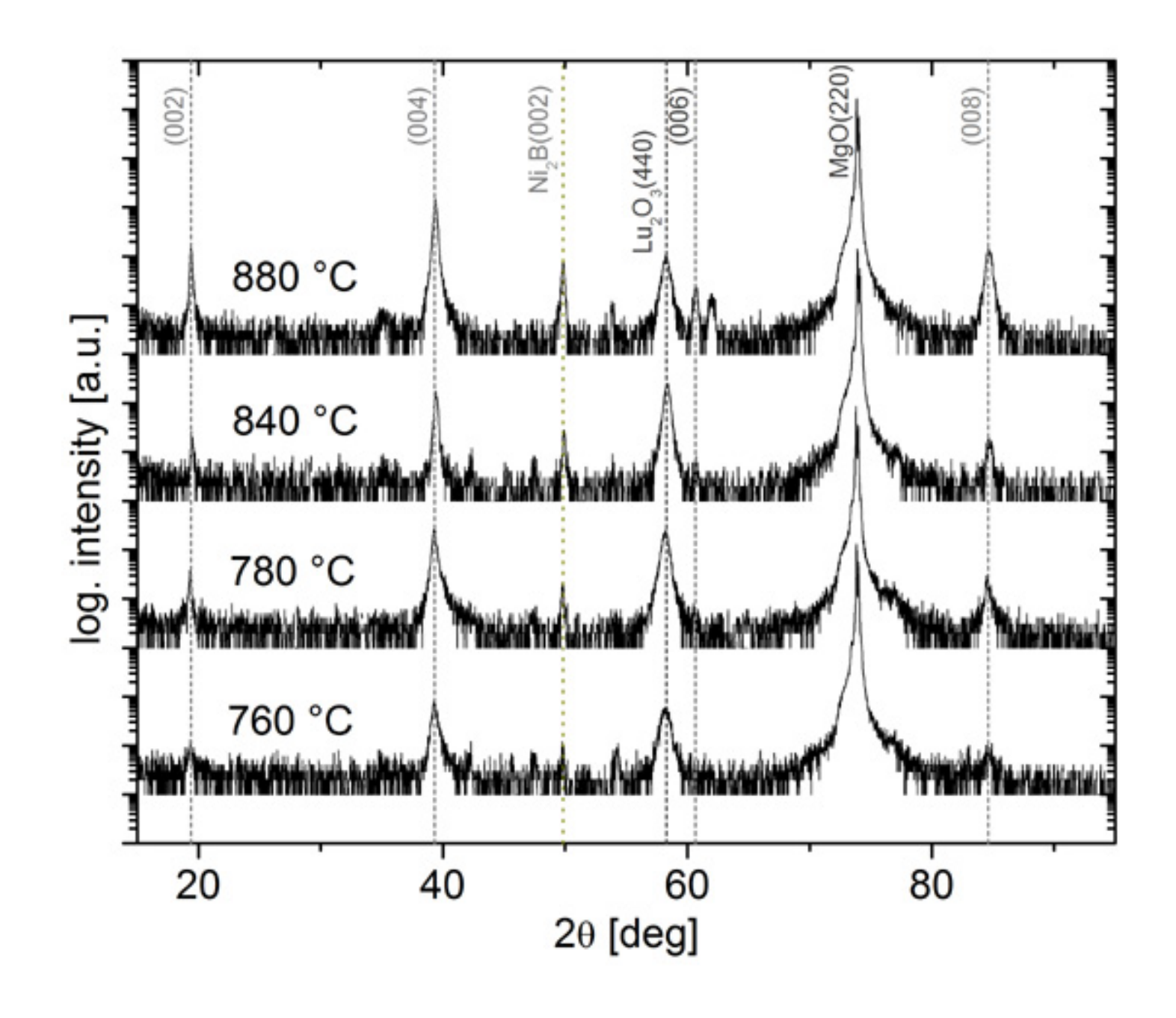}
\end{center}
\caption{X-ray measurements of \luni thin films deposited
onto MgO(110) substrates at different substrate temperatures (laser repetition rate $f$ = 20 Hz). The c--axis texture is almost perfect. The phase formation is clearly enhanced with increasing temperature while for the highest deposition temperature of 880 \textcelsius\/, an increasing amount of secondary phase formation is observed.}
\end{figure}

\begin{figure}[h]
\begin{center}
\includegraphics[width=7cm]{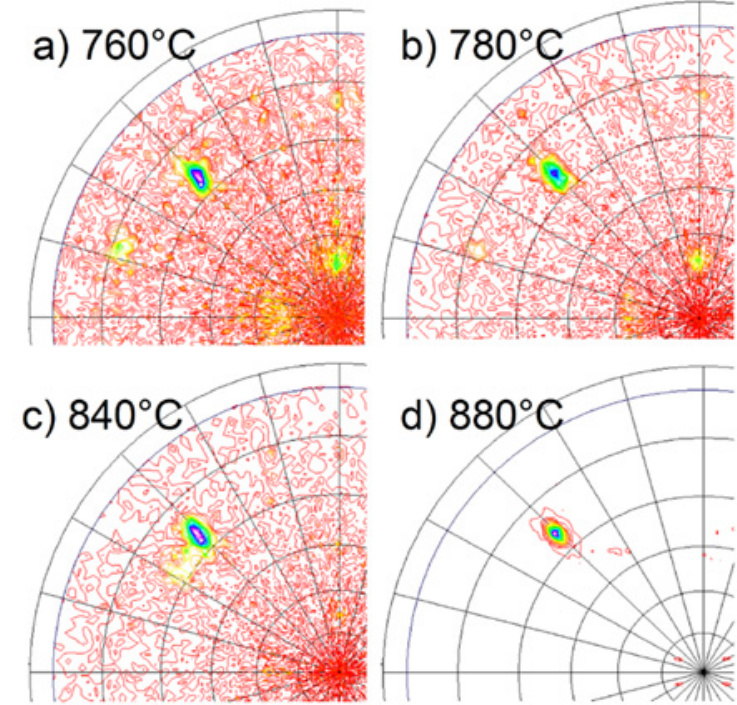}
\end{center}
\caption{(color online). Pole figure measurements of the \luni(112) reflection of the four films from Fig.\,2 at 2$\theta$ = 40.05\textdegree\/ (Cu K$_{\rm{\alpha}}$). The z--axis is scaled linearly and normalized for each measurement. Only one quadrant is shown due to the in--plane symmetry of the borocarbide crystal. The films exhibit almost perfect in--plane order.}
\end{figure}
Texture measurements were performed on the prepared samples in order to check the in--plane alignment of the grown films. The results are summarized in Fig.\,3. The (112) plane of \luni was chosen for these investigations as the highest intensity in powder diffraction is exhibited. Due to the tetragonal crystal structure with its double mirror symmetry of the basal plane, one quadrant of the pole figure already contains the necessary information.
\begin{figure}[h]
\begin{center}
\includegraphics[width=8cm]{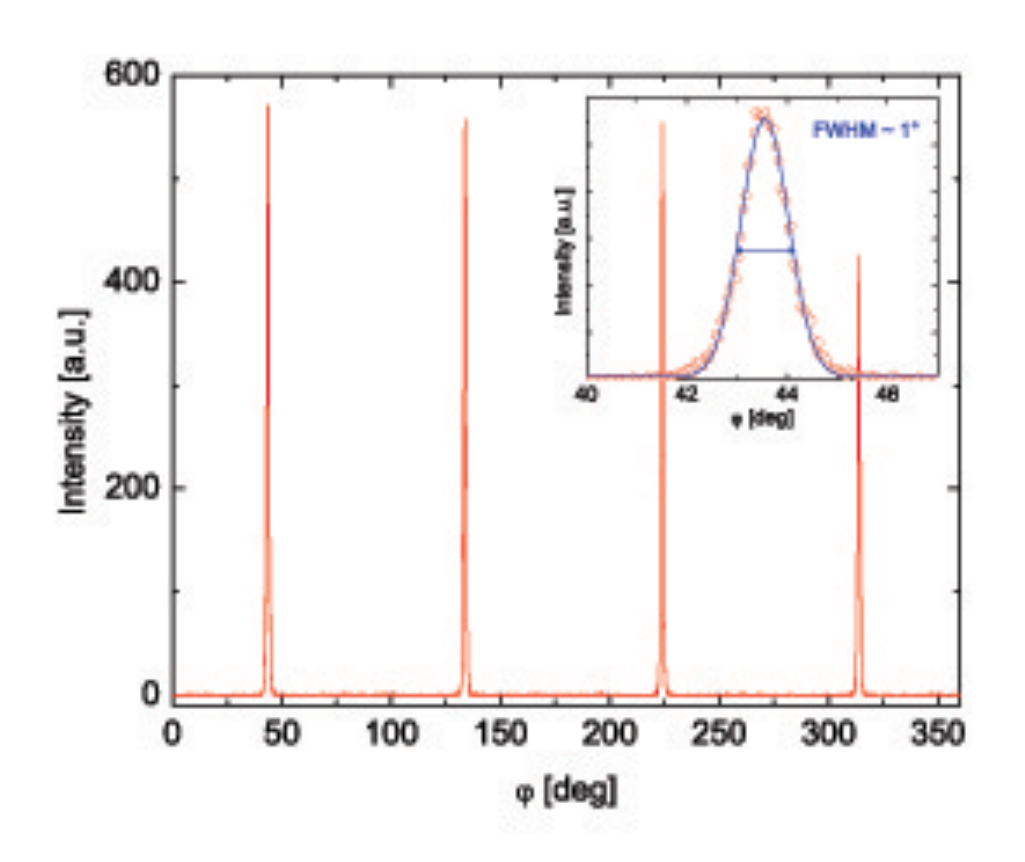}
\end{center}
\caption{(color online). X--ray phi scan of a \luni film on MgO(110) deposited at 5 Hz at 830 \textcelsius. The (112) reciprocal lattice plane at $\psi$ = 65\textdegree\/ was used for the measurement. The in--plane full width at half maximum (FWHM) is about 1\textdegree\/ (Gaussian fit in the inset).}
\end{figure}
A sharp peak at ($\phi$,$\psi$) = (45\textdegree, 65\textdegree) is observed in the films deposited at different temperatures. This indicates an epitaxial relationship between the substrate and the borocarbide in the following way:
\begin{equation}
 (110)[010]_{\mbox{\footnotesize MgO}} \; || \;  (010)[010]_{\mbox{\footnotesize \luni}}
\end{equation}
In agreement with the results of Fig.\,2, the intensity of the borocarbide (112) peak is strongly enhanced for higher deposition temperatures while a full width of half maximum (FWHM) of the phi scan of about 2\textdegree\/ is almost preserved. It should be noted further at this place that also the \luo interface layer has an epitaxial  relationship to the substrate, which can be described as
\begin{equation}
(110)[010]_{\mbox{\footnotesize MgO}} \; || \;
(110)[010]_{\mbox{\footnotesize \luo}}.
\end{equation}
A further improvement of the film quality was achieved using lower laser repetition rates. \luni films with the best out--of--plane and in--plane texture were prepared with a frequency of 5 Hz under slightly optimized conditions. In Fig.\,4, a phi scan of the (112) pole of such the sample is shown: The FWHM is about 1\textdegree\/ only, which is the lowest value reported so far on borocarbide thin films. The out--of--plane alignment has a FWHM value about 2.5 \textdegree\/ (not shown). Taking into account the low background signal, these data testify an extremely high in-plane order achieved under these optimized deposition conditions compared to earlier studies on borocarbide thin films.
\section{Superconductive properties }
The superconductive properties of the prepared \luni films were studied with resistive measurements in a \emph{Quantum Design PPMS}. The films were resistively characterized in the as--prepared state using a four-point resistivity measurement, where the four contacts were linearly aligned in a row and placed about 2 mm from the sample edge. The results of the temperature series prepared with a  laser repetition rate of 20 Hz are summarized in table 1.
\begin{figure}[h]
\begin{center}
\includegraphics[width=8.5cm]{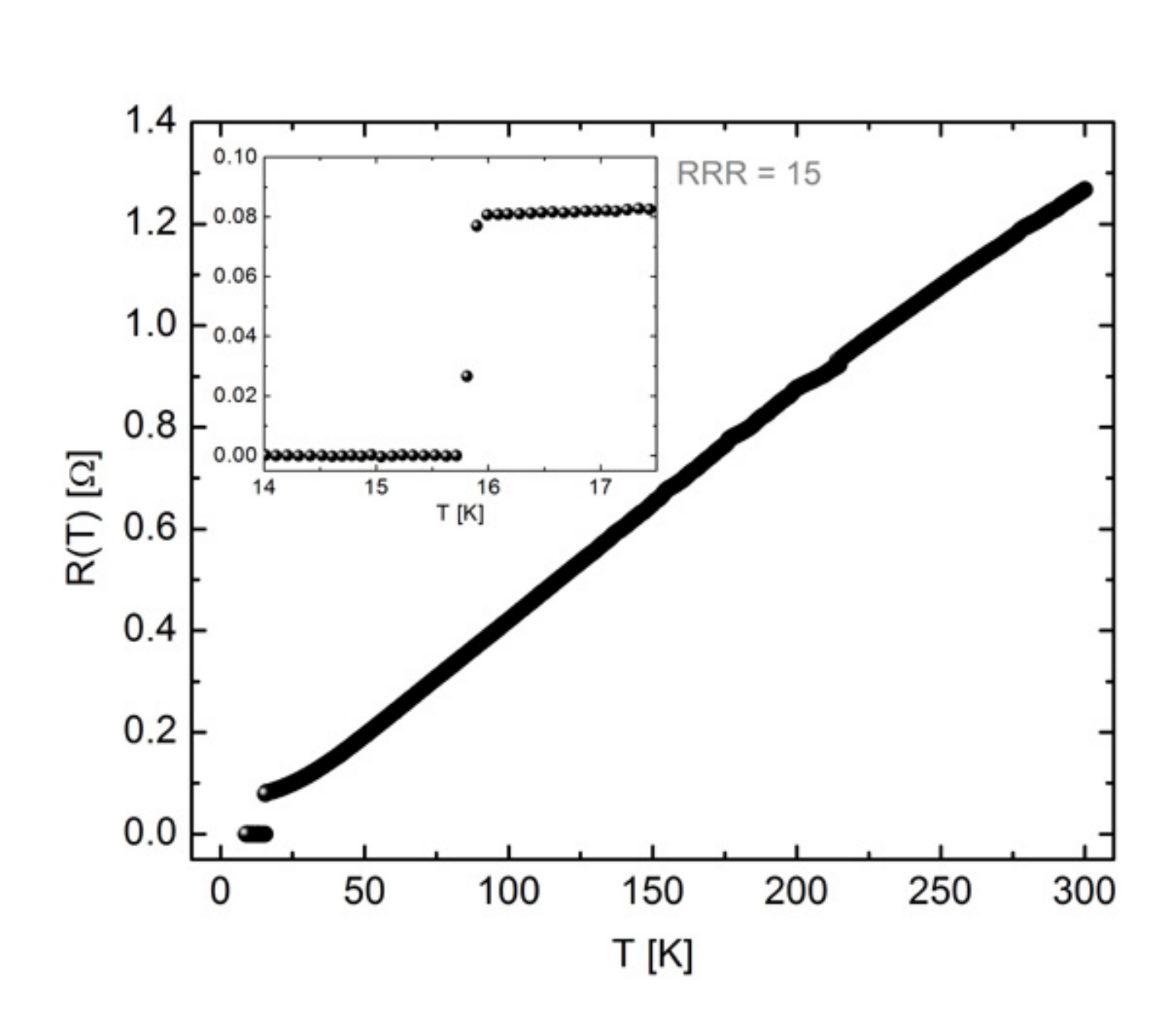}
\includegraphics[width=8.5cm]{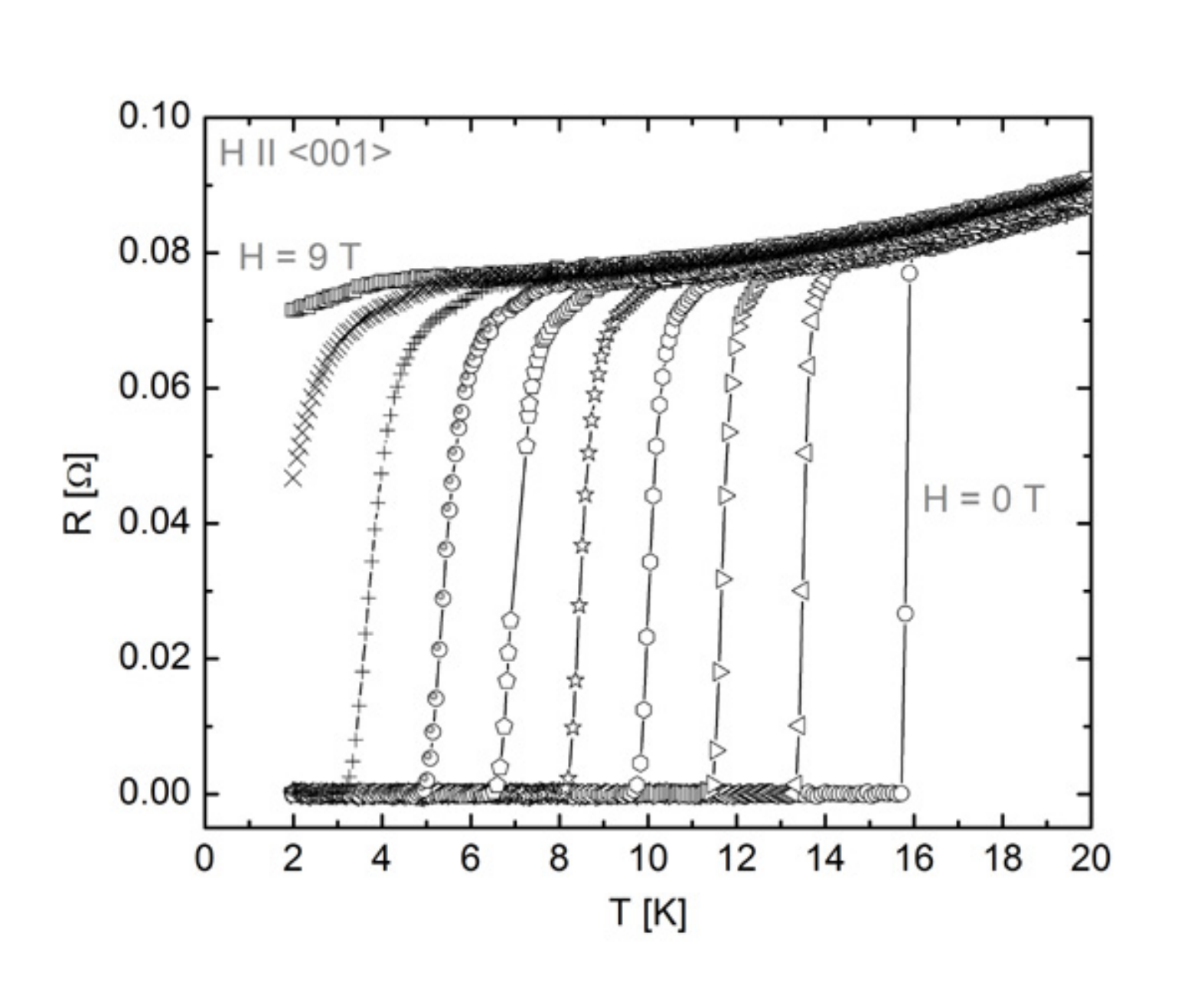}
\end{center}
\caption{Resistive measurements of a sample prepared at 5 Hz under optimized conditions. The data were achieved using a four-point probe at $T$/$t$ = 1 K/min (R($T$) from 2 to 300 K: 5 K/min) and at a measurement current $I$ = 1 mA. (a) R($T$) measurement in the absence of an external magnetic field exhibiting a residual resistivity ratio of about 15. The superconducting transition with a $\Delta$\tc\/ $\approx$ 0.3 K is shown in the inset. (b) R(T) at different external magnetic fields aligned parallel to the [001] borocarbide crystallographic axis.}
\end{figure}
It is apparent that the critical temperatures of these films with a value of 14 to 15.1 K are considerably below the bulk value of up to 16.5K. However, the increasing critical temperatures with increasing deposition temperature up to 840 \textcelsius\/ clearly point out that with improved phase formation and in--plane texture intensity, better superconducting properties are achieved. For a deposition temperature of 880\textdegree\/ however, the superconducting transition is dramatically broadened and the sample eventually becomes completely superconducting at a very low temperature of about 4 K only. We conclude that the superconducting phase is affected by increased interdiffusion of impurity phases from the interface into the film at this temperature. Thus, deposition temperatures around 840 \textcelsius\/ are kept during further optimization of the deposition conditions. Consequently, the \tc\/ values resistively determined from samples fabricated under optimized deposition conditions ($T$ = 830 \textcelsius, $f$ = 5 Hz) of up to 15.8 K are considerably closer to the optimum and typically sharp $\Delta$\tc\/ $<$ 0.3 K are observed.
\begin{figure}[h]
\includegraphics[width=8.5cm]{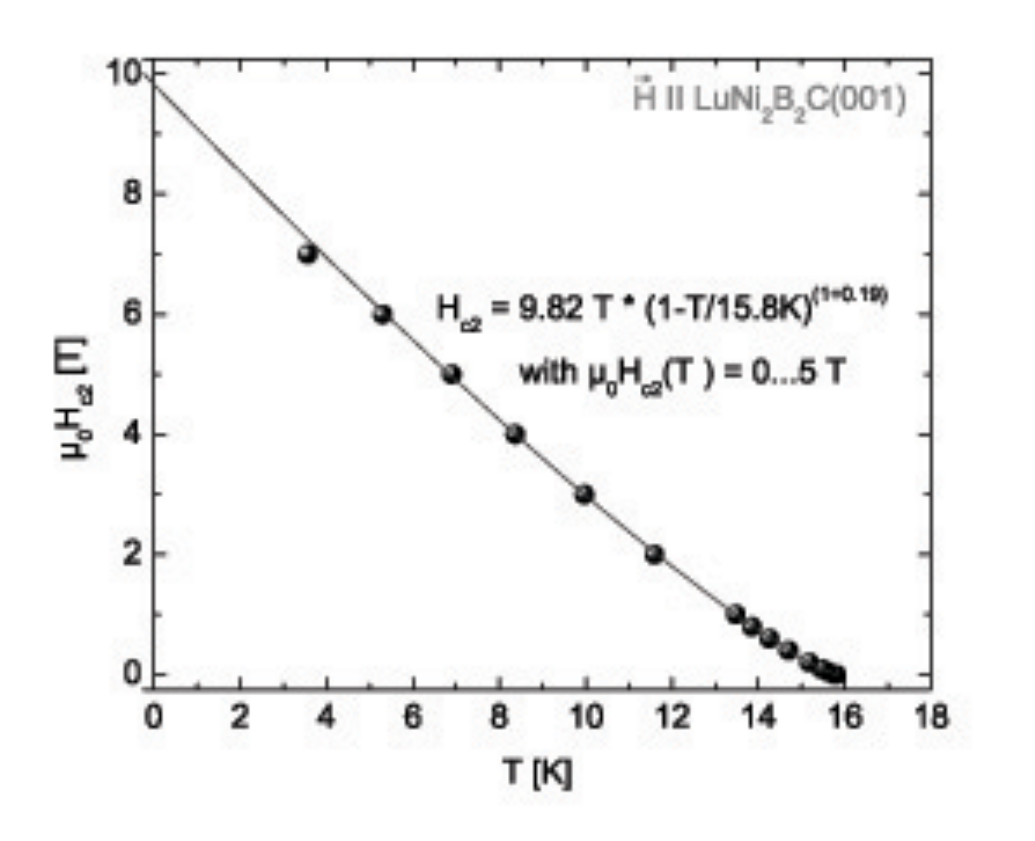}
\caption{Upper critical field of the \luni film in [001] direction, extracted from the midpoints of the R($T$) curves shown in Fig. 5. The data is fitted using the power law $H_{c2}(T) = H_{c2}(0) \cdot (1-T/T_c)^{(1+\alpha)}$. The positive curvature for $T \lessapprox$ \tc\/ calculated from the fit is about $\alpha$ = 0.19 which indicates a relatively low intraband scattering in the film.}
\end{figure}

\begin{table}[h]
\begin{tabular}{|c|c|c|c|c|}
  \hline
  T$_{\mbox{\scriptsize Dep}}$ [\textcelsius] & T$_{\mbox{\scriptsize
c}}^{\mbox{\scriptsize 90}}$ [K] & $\Delta$\tc\/ [K] \\
  \hline
  760 & 14.1 & 0.4 \\
  780 & 14.1 & 0.3 \\
  840 & 15.1 & 0.4 \\
  880 & 14.3 & 6.7 \\
  \hline
\end{tabular}
\caption{Onset of the resistive superconduction transition temperatures
(T$_{\mbox{\scriptsize c}}^{\mbox{\scriptsize 90}}$) and resistive transition widths ($\Delta$\tc) of \luni films prepared at different temperatures using a laser repetition rate of 20 Hz}
\end{table}
The results of the resistive measurements on the \luni layer deposited under optimized conditions ($T$ = 830 \textcelsius, $f$ = 5 Hz) are shown in Fig.\,5 (a) and (b). The external magnetic field $H$ was aligned parallel to the c--axis of the borocarbide which here is parallel to the normal axis of the films. The transition curves are quite steep as seen in Fig.\,5(b) and the slight broadening for higher magnetic fields, caused by fluctuations in the film, is known from crystalline samples as well \cite{MWK+97}. The residual resistivity ratio (RRR) of the film was estimated to a value of about 15 from the R($T$) curve which is close to the highest values reported for borocarbide thin films so far (Fig.\,5(a)). Thus, the temperature dependence of the upper critical field \hc\/ can be extracted from the R($T$) curves taking the temperature values of the midpoints of the R($T$) transition curves for instance. As a result, the upper critical field \hc($T$) for H $\parallel$ (001) is shown in Fig.\,6. It clearly shows a positive curvature close to the critical temperature in zero field, indicating a sufficiently low intraband scattering. Using a common power law fit $H_{c2}(T) = H_{c2}(0) \cdot (1-T/T_c)^{(1+\alpha)}$ for T $>$ 6 K, a zero temperature critical field of \hc(0) $\approx$ 9.82 T and an exponent $\alpha$ = 0.19 is calculated. This \hc(0) value extrapolated from that fit is close to the value of 9.3 T given in \cite{SDB+01} for a \luni single crystal with a critical temperature \tc $\approx$ 16.2 K and indicates a good agreement between the intrinsic properties of the thin film and the single crystal.
\section{Summary}
In summary, thin \luni film samples were prepared, which reproducibly combine almost perfect epitaxial growth with excellent superconductive properties (RRR up to 15, \tc\/ up to 15.8 K). The epitaxial growth is transferred from the MgO substrate to the borocarbide thin film via a perfectly textured lutetium oxide interface layer and can leads to a complete epitaxial in--plane order of the borocarbide film. The optimized films' electrical resistance shows a critical temperature close to the single crystal optimum and a transition width of about 0.3 K in the absence of an external magnetic field. This combination of excellent structural and superconducting properties was achieved for the first time in borocarbide thin films. The measurement of the upper critical field along the [001] crystallographic direction is highly comparable with exemplary single crystal data. Consequently, the films can be almost adequately used for investigations of intrinsic properties of the \luni phase.
\section{Acknowledgements}
The authors thank M. K\"uhnel and U. Besold for comprehensive laboratory services and K. Tscharntke for helpful thickness measurements. Helpful discussions with G. Fuchs are gratefully acknowledged. This work was financially supported by the Deutsche Forschungsgemeinschaft through the framework SFB 463.


\begin{thebibliography}{29}
\expandafter\ifx\csname natexlab\endcsname\relax\def\natexlab#1{#1}\fi
\expandafter\ifx\csname bibnamefont\endcsname\relax
  \def\bibnamefont#1{#1}\fi
\expandafter\ifx\csname bibfnamefont\endcsname\relax
  \def\bibfnamefont#1{#1}\fi
\expandafter\ifx\csname citenamefont\endcsname\relax
  \def\citenamefont#1{#1}\fi
\expandafter\ifx\csname url\endcsname\relax
  \def\url#1{\texttt{#1}}\fi
\expandafter\ifx\csname urlprefix\endcsname\relax\def\urlprefix{URL }\fi
\providecommand{\bibinfo}[2]{#2}
\providecommand{\eprint}[2][]{\url{#2}}

\bibitem[{\citenamefont{Cava et~al.}(1994{\natexlab{a}})\citenamefont{Cava,
  Zandbergen, Batlogg, Eisaki, Takagi, Krajewski, Peck, Gyorgy, and
  Uchida}}]{CZB+94}
\bibinfo{author}{\bibfnamefont{R.J.}~\bibnamefont{Cava}},
  \bibinfo{author}{\bibfnamefont{H.}~\bibnamefont{Zandbergen}},
  \bibinfo{author}{\bibfnamefont{B.}~\bibnamefont{Batlogg}},
  \bibinfo{author}{\bibfnamefont{H.}~\bibnamefont{Eisaki}},
  \bibinfo{author}{\bibfnamefont{H.}~\bibnamefont{Takagi}},
  \bibinfo{author}{\bibfnamefont{J.J.}~\bibnamefont{Krajewski}},
  \bibinfo{author}{\bibfnamefont{W.F.}~\bibnamefont{Peck}},
  \bibinfo{author}{\bibfnamefont{E.}~\bibnamefont{Gyorgy}}, \bibnamefont{and}
  \bibinfo{author}{\bibfnamefont{S.}~\bibnamefont{Uchida}},
  \bibinfo{journal}{Nature} \textbf{\bibinfo{volume}{372}},
  \bibinfo{pages}{245} (\bibinfo{year}{1994}{\natexlab{a}}).

\bibitem[{\citenamefont{Cava et~al.}(1994{\natexlab{b}})\citenamefont{Cava,
  Batlogg, Siegrist, Krajewski, Peck, Carter, Felder, Takagi, and
  vanDover}}]{CBS+94}
\bibinfo{author}{\bibfnamefont{R.J.}~\bibnamefont{Cava}},
  \bibinfo{author}{\bibfnamefont{B.}~\bibnamefont{Batlogg}},
  \bibinfo{author}{\bibfnamefont{T.}~\bibnamefont{Siegrist}},
  \bibinfo{author}{\bibfnamefont{J.J.}~\bibnamefont{Krajewski}},
  \bibinfo{author}{\bibfnamefont{W.F.}~\bibnamefont{Peck}},
  \bibinfo{author}{\bibfnamefont{S.}~\bibnamefont{Carter}},
  \bibinfo{author}{\bibfnamefont{R.J.}~\bibnamefont{Felder}},
  \bibinfo{author}{\bibfnamefont{H.}~\bibnamefont{Takagi}}, \bibnamefont{and}
  \bibinfo{author}{\bibfnamefont{R.B.}~\bibnamefont{vanDover}},
  \bibinfo{journal}{Phys.Rev.B} \textbf{\bibinfo{volume}{49}},
  \bibinfo{pages}{12384} (\bibinfo{year}{1994}{\natexlab{b}}).

\bibitem[{\citenamefont{Siegrist
  et~al.}(1994{\natexlab{a}})\citenamefont{Siegrist, Zandbergen, Cava,
  Krajewski, and Peck}}]{SZC+94}
\bibinfo{author}{\bibfnamefont{T.}~\bibnamefont{Siegrist}},
  \bibinfo{author}{\bibfnamefont{H.}~\bibnamefont{Zandbergen}},
  \bibinfo{author}{\bibfnamefont{R.J.}~\bibnamefont{Cava}},
  \bibinfo{author}{\bibfnamefont{J.}~\bibnamefont{Krajewski}},
  \bibnamefont{and} \bibinfo{author}{\bibfnamefont{W.F.}~\bibnamefont{Peck}},
  \bibinfo{journal}{Nature} \textbf{\bibinfo{volume}{367}},
  \bibinfo{pages}{254} (\bibinfo{year}{1994}{\natexlab{a}}).

\bibitem[{\citenamefont{Siegrist
  et~al.}(1994{\natexlab{b}})\citenamefont{Siegrist, Cava, Krajewski, and
  Peck}}]{SCKP94}
\bibinfo{author}{\bibfnamefont{T.}~\bibnamefont{Siegrist}},
  \bibinfo{author}{\bibfnamefont{R.J.}~\bibnamefont{Cava}},
  \bibinfo{author}{\bibfnamefont{J.}~\bibnamefont{Krajewski}},
  \bibnamefont{and} \bibinfo{author}{\bibfnamefont{W.F.}~\bibnamefont{Peck}},
  \bibinfo{journal}{J.Alloys Comp.} \textbf{\bibinfo{volume}{216}},
  \bibinfo{pages}{135} (\bibinfo{year}{1994}{\natexlab{b}}).

\bibitem[{\citenamefont{Muller and Narozhnyi}(2001)}]{MN01}
\bibinfo{author}{\bibfnamefont{K.}~\bibnamefont{Muller}} \bibnamefont{and}
  \bibinfo{author}{\bibfnamefont{V.}~\bibnamefont{Narozhnyi}},
  \bibinfo{journal}{Reports on Progress in Physics}
  \textbf{\bibinfo{volume}{64}}, \bibinfo{pages}{943} (\bibinfo{year}{2001}).

\bibitem[{\citenamefont{Thalmeier and Zwicknagl}(2005)}]{TZ05}
\bibinfo{author}{\bibfnamefont{P.}~\bibnamefont{Thalmeier}} \bibnamefont{and}
  \bibinfo{author}{\bibfnamefont{G.}~\bibnamefont{Zwicknagl}}, in
  \emph{\bibinfo{booktitle}{Handbook on the Physics and Chemistry of Rare
  Earths}}, edited by
  \bibinfo{editor}{\bibfnamefont{K.}~\bibnamefont{Gschneidner~Jr.}},
  \bibinfo{editor}{\bibfnamefont{J.-C.} \bibnamefont{B\"unzli}},
  \bibnamefont{and} \bibinfo{editor}{\bibfnamefont{V.}~\bibnamefont{Pecharsky}}
  (\bibinfo{publisher}{Elsevier B.V.}, \bibinfo{year}{2005}), vol.
  \bibinfo{volume}{34, chapt. 219}.

\bibitem[{\citenamefont{Mazumdar and Nagarajan}(2005)}]{MN05}
\bibinfo{author}{\bibfnamefont{C.}~\bibnamefont{Mazumdar}} \bibnamefont{and}
  \bibinfo{author}{\bibfnamefont{R.}~\bibnamefont{Nagarajan}},
  \bibinfo{journal}{Current Science} \textbf{\bibinfo{volume}{88}},
  \bibinfo{pages}{83} (\bibinfo{year}{2005}).
  
\bibitem[{\citenamefont{M\"uller et~al.}(2008)\citenamefont{M\"uller, Schneider, Fuchs and Drechsler}}]{MSFD08}
\bibinfo{author}{\bibfnamefont{K.H.}~\bibnamefont{M\"uller}},
  \bibinfo{author}{\bibfnamefont{M.}~\bibnamefont{Schneider}},
  \bibinfo{author}{\bibfnamefont{G.}~\bibnamefont{Fuchs}},
  \bibnamefont{and}
  \bibinfo{author}{\bibfnamefont{S.L.}~\bibnamefont{Drechsler}}, in
  \emph{\bibinfo{booktitle}{Handbook on the Physics and Chemistry of Rare
  Earths}}, edited by
  \bibinfo{editor}{\bibfnamefont{K.}~\bibnamefont{Gschneidner~Jr.}},
  \bibinfo{editor}{\bibfnamefont{J.-C.} \bibnamefont{B\"unzli}},
  \bibnamefont{and} \bibinfo{editor}{\bibfnamefont{V.}~\bibnamefont{Pecharsky}}
  (\bibinfo{publisher}{Elsevier B.V.}, \bibinfo{year}{2008}), vol.
  \bibinfo{volume}{38, chapt 239}.

\bibitem[{\citenamefont{Mattheiss}(1995)}]{Mat95}
\bibinfo{author}{\bibfnamefont{L.}~\bibnamefont{Mattheiss}},
  \bibinfo{journal}{Solid State Communications} \textbf{\bibinfo{volume}{94}},
  \bibinfo{pages}{741} (\bibinfo{year}{1995}).

\bibitem[{\citenamefont{Bud'ko and Canfield}(2006)}]{BC06}
\bibinfo{author}{\bibfnamefont{S.L.}~\bibnamefont{Bud'ko}} \bibnamefont{and}
  \bibinfo{author}{\bibfnamefont{P.C.}~\bibnamefont{Canfield}},
  \bibinfo{journal}{Comptes Rendus Physique} \textbf{\bibinfo{volume}{7}},
  \bibinfo{pages}{56} (\bibinfo{year}{2006}).

\bibitem[{\citenamefont{Behr et~al.}(2008)\citenamefont{Behr, Loser, Souptel,
  Fuchs, Mazilu, Cao, K\"ohler, Schultz, and B\"uchner}}]{BLS+08}
\bibinfo{author}{\bibfnamefont{G.}~\bibnamefont{Behr}},
  \bibinfo{author}{\bibfnamefont{W.}~\bibnamefont{L\"oser}},
  \bibinfo{author}{\bibfnamefont{D.}~\bibnamefont{Souptel}},
  \bibinfo{author}{\bibfnamefont{G.}~\bibnamefont{Fuchs}},
  \bibinfo{author}{\bibfnamefont{I.}~\bibnamefont{Mazilu}},
  \bibinfo{author}{\bibfnamefont{C.}~\bibnamefont{Cao}},
  \bibinfo{author}{\bibfnamefont{A.}~\bibnamefont{K\"ohler}},
  \bibinfo{author}{\bibfnamefont{L.}~\bibnamefont{Schultz}}, \bibnamefont{and}
  \bibinfo{author}{\bibfnamefont{B.}~\bibnamefont{B\"uchner}},
  \bibinfo{journal}{J.Cryst.Growth} \textbf{\bibinfo{volume}{310}},
  \bibinfo{pages}{2268} (\bibinfo{year}{2008}).

\bibitem[{\citenamefont{Souptel et~al.}(2005)\citenamefont{Souptel, Behr,
  Kreyssig, and L\"oser}}]{SBKL05}
\bibinfo{author}{\bibfnamefont{D.}~\bibnamefont{Souptel}},
  \bibinfo{author}{\bibfnamefont{G.}~\bibnamefont{Behr}},
  \bibinfo{author}{\bibfnamefont{A.}~\bibnamefont{Kreyssig}}, \bibnamefont{and}
  \bibinfo{author}{\bibfnamefont{W.}~\bibnamefont{L\"oser}},
  \bibinfo{journal}{J.Cryst.Growth} \textbf{\bibinfo{volume}{276}},
  \bibinfo{pages}{652} (\bibinfo{year}{2005}).

\bibitem[{\citenamefont{Bergk et~al.}(2007)\citenamefont{Bergk, Ignatchik,
  Bianchi, Jaeckel, Wosnitza, Perenboom, and Canfield}}]{BIB+07}
\bibinfo{author}{\bibfnamefont{B.}~\bibnamefont{Bergk}},
  \bibinfo{author}{\bibfnamefont{O.}~\bibnamefont{Ignatchik}},
  \bibinfo{author}{\bibfnamefont{A.}~\bibnamefont{Bianchi}},
  \bibinfo{author}{\bibfnamefont{M.}~\bibnamefont{Jaeckel}},
  \bibinfo{author}{\bibfnamefont{J.}~\bibnamefont{Wosnitza}},
  \bibinfo{author}{\bibfnamefont{J.}~\bibnamefont{Perenboom}},
  \bibnamefont{and} \bibinfo{author}{\bibfnamefont{P.}~\bibnamefont{Canfield}},
  \bibinfo{journal}{Physica C} \textbf{\bibinfo{volume}{460}},
  \bibinfo{pages}{630} (\bibinfo{year}{2007}).

\bibitem[{\citenamefont{Naidyuk et~al.}(2007)\citenamefont{Naidyuk, Bashlakov,
  Yanson, Fuchs, Behr, Souptel, and Drechsler}}]{NBY+07}
\bibinfo{author}{\bibfnamefont{Y.}~\bibnamefont{Naidyuk}},
  \bibinfo{author}{\bibfnamefont{D.}~\bibnamefont{Bashlakov}},
  \bibinfo{author}{\bibfnamefont{I.}~\bibnamefont{Yanson}},
  \bibinfo{author}{\bibfnamefont{G.}~\bibnamefont{Fuchs}},
  \bibinfo{author}{\bibfnamefont{G.}~\bibnamefont{Behr}},
  \bibinfo{author}{\bibfnamefont{D.}~\bibnamefont{Souptel}}, \bibnamefont{and}
  \bibinfo{author}{\bibfnamefont{S.}~\bibnamefont{Drechsler}},
  \bibinfo{journal}{Physica C} \textbf{\bibinfo{volume}{460}},
  \bibinfo{pages}{103} (\bibinfo{year}{2007}).

\bibitem[{\citenamefont{Bobrov et~al.}(2008)\citenamefont{Bobrov, Chernobay,
  Naidyuk, Tyutrina, Naugle, Rathnayaka, Bud'ko, Canfield, and
  Yanson}}]{BCN+08}
\bibinfo{author}{\bibfnamefont{N.}~\bibnamefont{Bobrov}},
  \bibinfo{author}{\bibfnamefont{V.}~\bibnamefont{Chernobay}},
  \bibinfo{author}{\bibfnamefont{Y.}~\bibnamefont{Naidyuk}},
  \bibinfo{author}{\bibfnamefont{L.}~\bibnamefont{Tyutrina}},
  \bibinfo{author}{\bibfnamefont{D.}~\bibnamefont{Naugle}},
  \bibinfo{author}{\bibfnamefont{K.}~\bibnamefont{Rathnayaka}},
  \bibinfo{author}{\bibfnamefont{S.}~\bibnamefont{Bud'ko}},
  \bibinfo{author}{\bibfnamefont{P.}~\bibnamefont{Canfield}}, \bibnamefont{and}
  \bibinfo{author}{\bibfnamefont{I.}~\bibnamefont{Yanson}},
  \bibinfo{journal}{Epl} \textbf{\bibinfo{volume}{83}},
  (\bibinfo{year}{2008}).

\bibitem[{\citenamefont{Schneider et~al.}(2008)\citenamefont{Schneider, Gladun,
  Kreyssig, Wosnitza, Petzold, Rosner, Behr, Souptel, M\"uller, Drechsler
  et~al.}}]{SGK+08}
\bibinfo{author}{\bibfnamefont{M.}~\bibnamefont{Schneider}},
  \bibinfo{author}{\bibfnamefont{A.}~\bibnamefont{Gladun}},
  \bibinfo{author}{\bibfnamefont{A.}~\bibnamefont{Kreyssig}},
  \bibinfo{author}{\bibfnamefont{J.}~\bibnamefont{Wosnitza}},
  \bibinfo{author}{\bibfnamefont{V.}~\bibnamefont{Petzold}},
  \bibinfo{author}{\bibfnamefont{H.}~\bibnamefont{Rosner}},
  \bibinfo{author}{\bibfnamefont{G.}~\bibnamefont{Behr}},
  \bibinfo{author}{\bibfnamefont{D.}~\bibnamefont{Souptel}},
  \bibinfo{author}{\bibfnamefont{K.}~\bibnamefont{M\"uller}},
  \bibinfo{author}{\bibfnamefont{S.}~\bibnamefont{Drechsler}},
  \bibnamefont{et~al.}, \bibinfo{journal}{J.Phys.: Cond.Matter}
  \textbf{\bibinfo{volume}{20}},  (\bibinfo{year}{2008}).

\bibitem[{\citenamefont{Arisawa et~al.}(1998)\citenamefont{Arisawa, Hatano,
  Nakamura, and Togano}}]{AHNT98}
\bibinfo{author}{\bibfnamefont{S.}~\bibnamefont{Arisawa}},
  \bibinfo{author}{\bibfnamefont{T.}~\bibnamefont{Hatano}},
  \bibinfo{author}{\bibfnamefont{K.}~\bibnamefont{Nakamura}}, \bibnamefont{and}
  \bibinfo{author}{\bibfnamefont{K.}~\bibnamefont{Togano}},
  \bibinfo{journal}{Physica C} \textbf{\bibinfo{volume}{308}},
  \bibinfo{pages}{67} (\bibinfo{year}{1998}).

\bibitem[{\citenamefont{Andreone et~al.}(1997)\citenamefont{Andreone, Aruta,
  Canepa, Cimberle, Cogliati, Ferdeghini, Fontana, Giannini, Guasconi, Iavarone
  et~al.}}]{AAC+97}
\bibinfo{author}{\bibfnamefont{A.}~\bibnamefont{Andreone}},
  \bibinfo{author}{\bibfnamefont{C.}~\bibnamefont{Aruta}},
  \bibinfo{author}{\bibfnamefont{F.}~\bibnamefont{Canepa}},
  \bibinfo{author}{\bibfnamefont{M.}~\bibnamefont{Cimberle}},
  \bibinfo{author}{\bibfnamefont{E.}~\bibnamefont{Cogliati}},
  \bibinfo{author}{\bibfnamefont{C.}~\bibnamefont{Ferdeghini}},
  \bibinfo{author}{\bibfnamefont{F.}~\bibnamefont{Fontana}},
  \bibinfo{author}{\bibfnamefont{E.}~\bibnamefont{Giannini}},
  \bibinfo{author}{\bibfnamefont{P.}~\bibnamefont{Guasconi}},
  \bibinfo{author}{\bibfnamefont{M.}~\bibnamefont{Iavarone}},
  \bibnamefont{et~al.}, \bibinfo{journal}{Nuovo Cimento Della Societa Italiana
  di Fisica D-Condensed Matter Atomic Molecular and Chemical Physics Fluids
  Plasmas Biophysics} \textbf{\bibinfo{volume}{19}}, \bibinfo{pages}{995}
  (\bibinfo{year}{1997}).

\bibitem[{\citenamefont{Andreone et~al.}(2001)\citenamefont{Andreone,
  Cassinese, Gianni, Iavarone, Palomba, and Vaglio}}]{ACG+01}
\bibinfo{author}{\bibfnamefont{A.}~\bibnamefont{Andreone}},
  \bibinfo{author}{\bibfnamefont{A.}~\bibnamefont{Cassinese}},
  \bibinfo{author}{\bibfnamefont{L.}~\bibnamefont{Gianni}},
  \bibinfo{author}{\bibfnamefont{M.}~\bibnamefont{Iavarone}},
  \bibinfo{author}{\bibfnamefont{F.}~\bibnamefont{Palomba}}, \bibnamefont{and}
  \bibinfo{author}{\bibfnamefont{R.}~\bibnamefont{Vaglio}},
  \bibinfo{journal}{Phys.Rev.B} \textbf{\bibinfo{volume}{6410}},
  (\bibinfo{year}{2001}).

\bibitem[{\citenamefont{Grassano et~al.}(2001)\citenamefont{Grassano, Marre,
  Pallecchi, Ricci, Siri, and Ferdeghini}}]{GMP+01}
\bibinfo{author}{\bibfnamefont{G.}~\bibnamefont{Grassano}},
  \bibinfo{author}{\bibfnamefont{D.}~\bibnamefont{Marre}},
  \bibinfo{author}{\bibfnamefont{I.}~\bibnamefont{Pallecchi}},
  \bibinfo{author}{\bibfnamefont{F.}~\bibnamefont{Ricci}},
  \bibinfo{author}{\bibfnamefont{A.}~\bibnamefont{Siri}}, \bibnamefont{and}
  \bibinfo{author}{\bibfnamefont{C.}~\bibnamefont{Ferdeghini}},
  \bibinfo{journal}{Supercond.Sci.Technol.} \textbf{\bibinfo{volume}{14}},
  \bibinfo{pages}{117} (\bibinfo{year}{2001}).

\bibitem[{\citenamefont{Haese et~al.}(2000)\citenamefont{Haese, Hough,
  Holzapfel, and Schultz}}]{HHHS00}
\bibinfo{author}{\bibfnamefont{K.}~\bibnamefont{Haese}},
  \bibinfo{author}{\bibfnamefont{D.}~\bibnamefont{Hough}},
  \bibinfo{author}{\bibfnamefont{B.}~\bibnamefont{Holzapfel}},
  \bibnamefont{and} \bibinfo{author}{\bibfnamefont{L.}~\bibnamefont{Schultz}},
  \bibinfo{journal}{Physica B} \textbf{\bibinfo{volume}{284}},
  \bibinfo{pages}{1105} (\bibinfo{year}{2000}).

\bibitem[{\citenamefont{Wimbush et~al.}(2001)\citenamefont{Wimbush, H\"ase,
  Schultz, and Holzapfel}}]{WHSH01}
\bibinfo{author}{\bibfnamefont{S.}~\bibnamefont{Wimbush}},
  \bibinfo{author}{\bibfnamefont{K.}~\bibnamefont{H\"ase}},
  \bibinfo{author}{\bibfnamefont{L.}~\bibnamefont{Schultz}}, \bibnamefont{and}
  \bibinfo{author}{\bibfnamefont{B.}~\bibnamefont{Holzapfel}},
  \bibinfo{journal}{J.Phys.: Cond.Matter} \textbf{\bibinfo{volume}{13}},
  \bibinfo{pages}{L355} (\bibinfo{year}{2001}).

\bibitem[{\citenamefont{Wimbush et~al.}(2003)\citenamefont{Wimbush, Schultz,
  and Holzapfel}}]{WSH03}
\bibinfo{author}{\bibfnamefont{S.}~\bibnamefont{Wimbush}},
  \bibinfo{author}{\bibfnamefont{L.}~\bibnamefont{Schultz}}, \bibnamefont{and}
  \bibinfo{author}{\bibfnamefont{B.}~\bibnamefont{Holzapfel}},
  \bibinfo{journal}{Physica C} \textbf{\bibinfo{volume}{388}},
  \bibinfo{pages}{191} (\bibinfo{year}{2003}).

\bibitem[{\citenamefont{Wimbush et~al.}(2004)\citenamefont{Wimbush, Schultz,
  and Holzapfel}}]{WSH04}
\bibinfo{author}{\bibfnamefont{S.}~\bibnamefont{Wimbush}},
  \bibinfo{author}{\bibfnamefont{L.}~\bibnamefont{Schultz}}, \bibnamefont{and}
  \bibinfo{author}{\bibfnamefont{B.}~\bibnamefont{Holzapfel}},
  \bibinfo{journal}{Physica C} \textbf{\bibinfo{volume}{408-10}},
  \bibinfo{pages}{83} (\bibinfo{year}{2004}).

\bibitem[{\citenamefont{Bashlakov et~al.}(2005)\citenamefont{Bashlakov,
  Naidyuk, Yanson, Wimbush, Holzapfel, Fuchs, and Drechsler}}]{BNY+05}
\bibinfo{author}{\bibfnamefont{D.}~\bibnamefont{Bashlakov}},
  \bibinfo{author}{\bibfnamefont{Y.}~\bibnamefont{Naidyuk}},
  \bibinfo{author}{\bibfnamefont{I.}~\bibnamefont{Yanson}},
  \bibinfo{author}{\bibfnamefont{S.}~\bibnamefont{Wimbush}},
  \bibinfo{author}{\bibfnamefont{B.}~\bibnamefont{Holzapfel}},
  \bibinfo{author}{\bibfnamefont{G.}~\bibnamefont{Fuchs}}, \bibnamefont{and}
  \bibinfo{author}{\bibfnamefont{S.}~\bibnamefont{Drechsler}},
  \bibinfo{journal}{Superconductor Science \& Technology}
  \textbf{\bibinfo{volume}{18,8}}, \bibinfo{pages}{1094}
  (\bibinfo{year}{2005}).

\bibitem[{\citenamefont{Arisawa et~al.}(1994)\citenamefont{Arisawa, Hatano,
  Hirata, Mochiku, Kitaguchi, Fujii, Kumakura, Kadowaki, Nakamura, and
  Togano}}]{AHH+94}
\bibinfo{author}{\bibfnamefont{S.}~\bibnamefont{Arisawa}},
  \bibinfo{author}{\bibfnamefont{T.}~\bibnamefont{Hatano}},
  \bibinfo{author}{\bibfnamefont{K.}~\bibnamefont{Hirata}},
  \bibinfo{author}{\bibfnamefont{T.}~\bibnamefont{Mochiku}},
  \bibinfo{author}{\bibfnamefont{H.}~\bibnamefont{Kitaguchi}},
  \bibinfo{author}{\bibfnamefont{H.}~\bibnamefont{Fujii}},
  \bibinfo{author}{\bibfnamefont{H.}~\bibnamefont{Kumakura}},
  \bibinfo{author}{\bibfnamefont{K.}~\bibnamefont{Kadowaki}},
  \bibinfo{author}{\bibfnamefont{K.}~\bibnamefont{Nakamura}}, \bibnamefont{and}
  \bibinfo{author}{\bibfnamefont{K.}~\bibnamefont{Togano}},
  \bibinfo{journal}{Appl.Phys.Lett.} \textbf{\bibinfo{volume}{65}},
  \bibinfo{pages}{1299} (\bibinfo{year}{1994}).

\bibitem[{\citenamefont{Cao et~al.}(2004)\citenamefont{Cao, Simon, and
  Skrotzki}}]{CSS04}
\bibinfo{author}{\bibfnamefont{G.}~\bibnamefont{Cao}},
  \bibinfo{author}{\bibfnamefont{P.}~\bibnamefont{Simon}}, \bibnamefont{and}
  \bibinfo{author}{\bibfnamefont{W.}~\bibnamefont{Skrotzki}},
  \bibinfo{journal}{J.Mater.Res.} \textbf{\bibinfo{volume}{19}},
  \bibinfo{pages}{1413} (\bibinfo{year}{2004}).

\bibitem[{\citenamefont{Reibold et~al.}(2002)\citenamefont{Reibold, Wimbush,
  Holzapfel, and Kramer}}]{RWHK02}
\bibinfo{author}{\bibfnamefont{M.}~\bibnamefont{Reibold}},
  \bibinfo{author}{\bibfnamefont{S.}~\bibnamefont{Wimbush}},
  \bibinfo{author}{\bibfnamefont{B.}~\bibnamefont{Holzapfel}},
  \bibnamefont{and} \bibinfo{author}{\bibfnamefont{U.}~\bibnamefont{Kramer}},
  \bibinfo{journal}{J.Alloys Comp.} \textbf{\bibinfo{volume}{347}},
  \bibinfo{pages}{24} (\bibinfo{year}{2002}).

\bibitem[{\citenamefont{Metlushko et~al.}(1997)\citenamefont{Metlushko, Welp,
  Koshelev, Aranson, Crabtree, and Canfield}}]{MWK+97}
\bibinfo{author}{\bibfnamefont{V.}~\bibnamefont{Metlushko}},
  \bibinfo{author}{\bibfnamefont{U.}~\bibnamefont{Welp}},
  \bibinfo{author}{\bibfnamefont{A.}~\bibnamefont{Koshelev}},
  \bibinfo{author}{\bibfnamefont{I.}~\bibnamefont{Aranson}},
  \bibinfo{author}{\bibfnamefont{G.W.}~\bibnamefont{Crabtree}}, \bibnamefont{and}
  \bibinfo{author}{\bibfnamefont{P.C.}~\bibnamefont{Canfield}},
  \bibinfo{journal}{Phys.Rev.Lett.} \textbf{\bibinfo{volume}{79}},
  \bibinfo{pages}{1738} (\bibinfo{year}{1997}).

\bibitem[{\citenamefont{Ferdeghini et~al.}(2003)\citenamefont{Ferdeghini,
  Grassano, Bellingeri, Marre, Ramadan, and Ferrando}}]{FGB+03}
\bibinfo{author}{\bibfnamefont{C.}~\bibnamefont{Ferdeghini}},
  \bibinfo{author}{\bibfnamefont{G.}~\bibnamefont{Grassano}},
  \bibinfo{author}{\bibfnamefont{E.}~\bibnamefont{Bellingeri}},
  \bibinfo{author}{\bibfnamefont{D.}~\bibnamefont{Marre}},
  \bibinfo{author}{\bibfnamefont{W.}~\bibnamefont{Ramadan}}, \bibnamefont{and}
  \bibinfo{author}{\bibfnamefont{V.}~\bibnamefont{Ferrando}},
  \bibinfo{journal}{Int.J.Mod.Phys.B} \textbf{\bibinfo{volume}{17}},
  \bibinfo{pages}{824} (\bibinfo{year}{2003}).

\bibitem[{\citenamefont{Schmiedeshoff et~al.}(2001)\citenamefont{Schmiedeshoff,
  Detwiler, Beyermann, Lacerda, Canfield, and Smith}}]{SDB+01}
\bibinfo{author}{\bibfnamefont{G.M.}~\bibnamefont{Schmiedeshoff}},
  \bibinfo{author}{\bibfnamefont{J.A.}~\bibnamefont{Detwiler}},
  \bibinfo{author}{\bibfnamefont{W.P.}~\bibnamefont{Beyermann}},
  \bibinfo{author}{\bibfnamefont{A.H.}~\bibnamefont{Lacerda}},
  \bibinfo{author}{\bibfnamefont{P.C.}~\bibnamefont{Canfield}}, \bibnamefont{and}
  \bibinfo{author}{\bibfnamefont{J.L.}~\bibnamefont{Smith}},
  \bibinfo{journal}{Phys.Rev.B} \textbf{\bibinfo{volume}{63}},
  (\bibinfo{year}{2001}).


\end{thebibliography}
\end{document}